\documentclass[twocolumn,superscriptaddress]{revtex4}
\usepackage{graphicx}
\usepackage{amsmath}
\usepackage{color}

\begin{document}

\title{Controlling Secondary Structures of Bio-Polymers with Hydrogen-Like Bonding}

\author{J.\ Krawczyk} \email{j.krawczyk@ms.unimelb.edu.au}
\affiliation{Department of Mathematics and Statistics, The University
  of Melbourne, 3010, Australia}
\author{A.\ L.\ Owczarek} \email{a.owczarek@ms.unimelb.edu.au}
\affiliation{Department of Mathematics and Statistics, The University
  of Melbourne, 3010, Australia}
\author{T.\ Prellberg} \email{t.prellberg@qmul.ac.uk}
\affiliation{School of Mathematical Sciences, Queen Mary, University
  of London, Mile End Road, London E1 4NS, United Kingdom}
\author{A.\ Rechnitzer} \email{andrewr@math.ubc.ca}
\affiliation{Department of Mathematics, University of British Columbia,
Vancouver, BC V6T 1Z2, Canada}

\begin{abstract}
  We present results for a lattice model of bio-polymers where the type of
  $\beta$-sheet formation can be controlled by different types of hydrogen
  bonds depending on the relative orientation of close segments of the
  polymer. Tuning these different interaction strengths leads to
  low-temperature structures with different types of orientational order.
  We perform simulations of this model and so present the phase diagram, ascertaining
  the nature of the phases and the order of the transitions between these phases.
\end{abstract}
\maketitle

\section{Introduction}

The transition of a flexible macromolecular chain from a random-coil
conformation to a globular compact form, called coil-globule
transition, has been a subject of extensive theoretical and
experimental studies \cite{baysal2003}.  Generally, polymers in a good
solvent are modelled by random walks with short-range repulsion
(excluded volume). Polymers undergoing a coil-globule transition are
then modelled by adding an additional short-range attraction. This
short-range attraction is both due to an affinity between monomers and
solvent molecules, affecting the solvability of a polymer, and also
due to intra-molecular interactions between different monomers, for
example due to van-der-Waals forces.
The canonical lattice model \cite{orr1946,bennett1998} for this
transition is given by interacting self-avoiding walks, in which
self-avoiding random walks on a lattice are weighted according to the
number of nearest-neighbour contacts (non-consecutively visited nearest-neighbour
lattice sites).

It is recognised that in biological systems, {\it e.g.} proteins, the most
relevant contribution to monomer-monomer-interactions is due to
hydrogen bonds. These hydrogen bonds can only form if neighbouring
segments are aligned in a certain way, resulting in an interaction
that is strongly dependent on the relative orientation of segments.
This type interaction plays a leading role in the formation of
secondary protein structures such as $\alpha$-helices and
$\beta$-sheets \cite{pauling1951}. In this paper, we introduce a model
for controlling the type of orientational order of these structures.

In \cite{bascle1993}, Bascle {\it et al} introduced a lattice model of
polymers interacting via hydrogen bonds, in which hydrogen bonds were
mimicked by an interaction between two nearest-neighbour lattice sites
which belong to two {\em straight} segments of the polymer.
This was treated in the context of
Hamiltonian walks in a mean-field approach, and they predicted a
first-order transition between an anisotropic ordered phase and a
molten phase.
Later, Foster and Seno introduced this type of interaction to a 
model of self-avoiding walks \cite{foster2001}. They analyzed it
using transfer-matrix techniques in two dimensions, where a first-order 
transition between a folded polymer crystal and a swollen coil was found.
Subsequently, a variant of this model was introduced by Buzano and
Pretti \cite{buzano2002}, where the interaction is defined between
parallel nearest-neighbour bonds, independent of
the straightness required in \cite{bascle1993}, arguing that these should
better take into account the contribution of fluctuating bonds, which may be
formed even in relatively disordered configurations. The
authors studied this interacting-bond model and the one
introduced by Foster and Seno on the square and simple cubic
lattice using the Bethe approximation.  They found a first-order
transition in the Foster-Seno model in two and three dimensions, confirming and
extending results in \cite{foster2001}. In contrast to this, they
found two transitions in the interacting-bond model, a second-order
$\theta$-transition from a swollen coil to a collapsed molten globule
and then a first-order transition to a folded polymer crystal.
In a later paper \cite{buzano2003}, they introduced a competing
isotropic interaction and studied its effect in three dimensions using
the Bethe approximation. They found a phase diagram with three
different phases (swollen coil, collapsed molten globule, folded polymer crystal),
similar to that of collapsing semi-stiff polymers
\cite{bastolla1997}.

In this work we generalize the Foster-Seno model to
distinguish between nearest-neighbour contacts of parallel and orthogonal 
straight segments (see Figure~\ref{interactions}) and
assign interactions of different strengths to these two types of
contacts, investigating it with Monte-Carlo simulations using the
the FlatPERM algorithm \cite{prellberg2004}.
We begin by simulating the Foster-Seno model and confirm
the theoretical picture presented above \cite{foster2001,buzano2003}.
We then consider our extended model (in three dimensions).
We find evidence for
two differently structured folded phases, depending on whether the
parallel or orthogonal interactions dominate. The transition between
the swollen coil and each of the two collapsed ordered crystals is first-order.
We investigate the structure of these two low-temperature phases. 
For strong parallel interactions long segments of the polymer align,
whereas for strong orthogonal
interactions the polymer forms alternating orthogonally layered $\beta$-sheets.

\section{Model and simulations}

\begin{figure}[h]
\center{
\includegraphics[scale=0.35,angle=0]{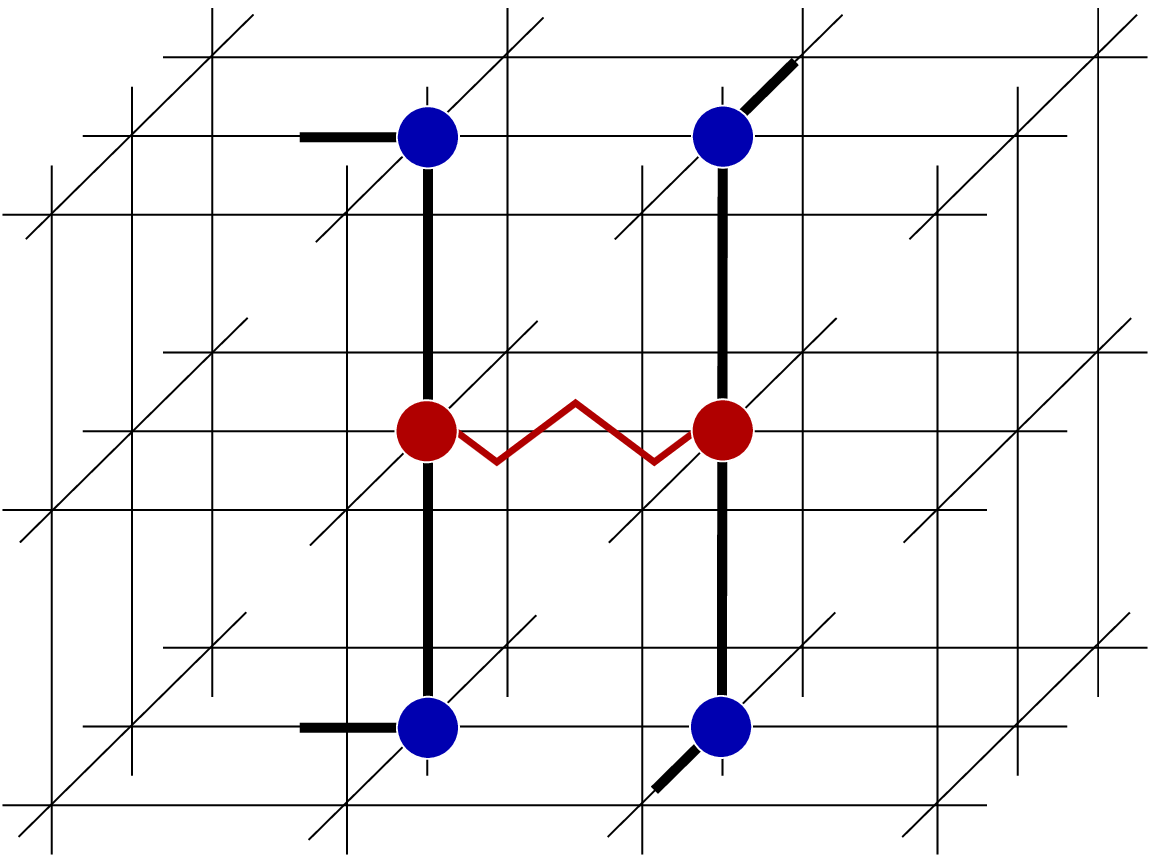}
\hfill
\includegraphics[scale=0.35,angle=0]{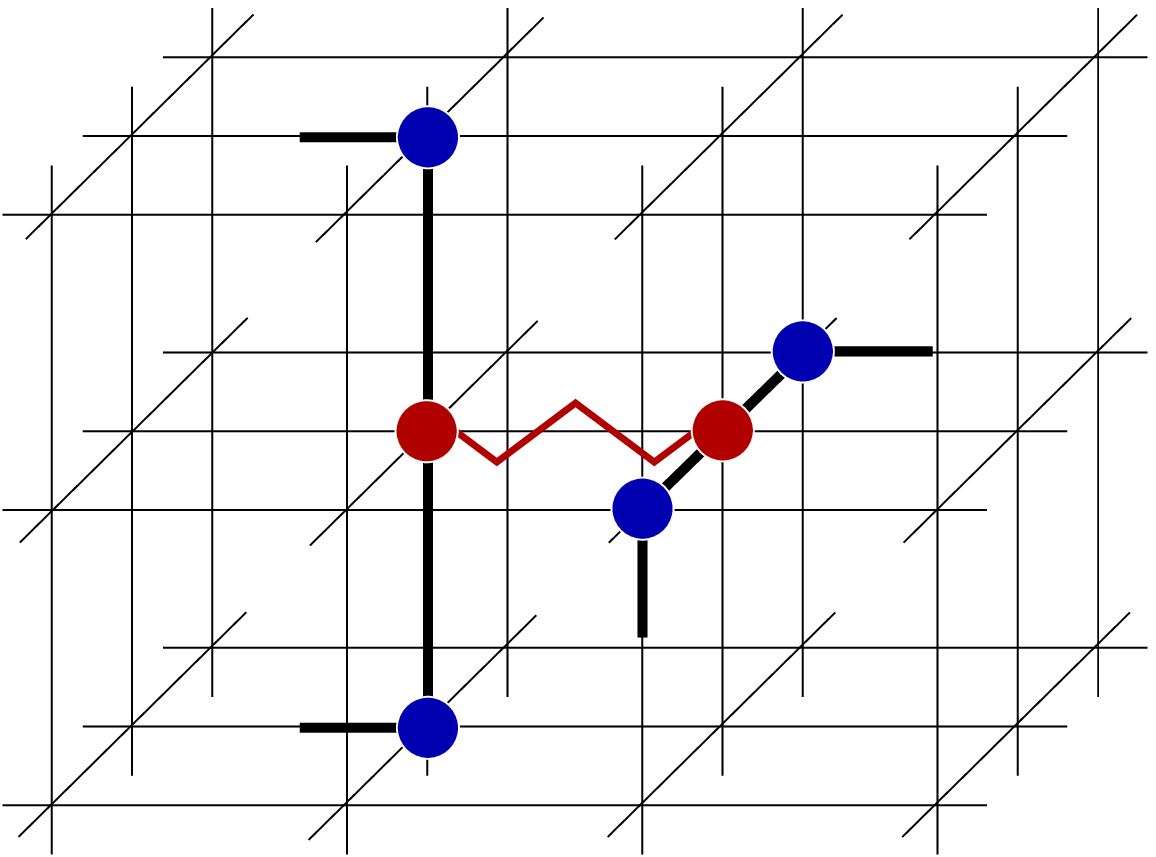}
\\
\hfill
parallel ({\bf p})
\hspace{3cm}
orthogonal ({\bf o})
\hfill
}
\caption{\label{interactions} The two types of nearest-neighbour 
  contacts between two straight segments of the polymer: parallel
  segments (left) with interaction $-\varepsilon_p$, and
  orthogonal segments (right) with interaction $-\varepsilon_o$.
  In two dimensions, only parallel interactions are
  possible.}
\end{figure}

A polymer is modelled as an $n$-step self-avoiding walk on the simple cubic lattice
with interactions $-\varepsilon_p$ and
$-\varepsilon_o$ for nearest-neighbour contacts
between parallel and orthogonal {\em straight} segments 
of the walk, as shown in Figure \ref{interactions}. Here, a segment
is defined as a site along with the two adjoining bonds visited by the walk,
and we say that a segment is straight if these two bonds are aligned.
The restriction of this model to $\varepsilon_o=\varepsilon_p$ is the simple generalisation of
the Foster-Seno model, which was originally defined on a square lattice, to three dimensions.

The total energy for a polymer configuration $\varphi_n$ with $n+1$
monomers (occupied lattice sites) is given by
\begin{equation}
E_n(\varphi_n) = - m_p(\varphi_n) \varepsilon_p - m_o(\varphi_n)\varepsilon_o
\end{equation}
depending on the number of non-consecutive parallel and orthogonal straight 
nearest-neighbour segments $m_p$ and $m_o$, respectively, along the polymer.
For convenience, we define
\begin{equation} 
\beta_p= \beta \varepsilon_p \mbox{ and } \beta_o= \beta \varepsilon_o \mbox{,}
\end{equation} 
where $\beta=1/k_BT$ for temperature $T$ and
Boltzmann constant $k_B$.  The partition function is given by
\begin{equation}
Z_n(\beta_p,\beta_o)=
\sum_{m_p,m_o} C_{n,m_p,m_o}\; e^{\beta_p m_p+\beta_o m_o}
\end{equation}
with $C_{n,m_p,m_o}$ being the density of states. 
We have simulated this model using the FlatPERM algorithm
\cite{prellberg2004}.  The power of this algorithm is the ability to
sample the density of states uniformly with respect to a chosen
parametrisation, so that the whole parameter range is accessible from
one simulation. In practice, we have also performed multiple independent
simulations to further reduce errors.
The natural parameters for this problem are $m_p$ and $m_o$, and the
algorithm directly estimates the density of states $C_{n,m_p,m_o}$ for
all $n\le n_{max}$ for some fixed $n_{max}$ and all possible values of $m_p$ and $m_o$.
Canonical averages are performed with respect to this density of states.
As we need to store the full density of
states, we only perform simulations up to a maximal length of
$n_{max}=128$, due to a memory requirement growing as $n^3$.
To reduce the error, we have taken
averages of ten independent runs each. Each run has taken
approximately 3 months on a 2.8GHz PC to complete.

Fixing one of the parameters $\beta_p$ and $\beta_o$ reduces the size
the histogram, and enables us to perform simulations of larger systems,
as the memory requirement now grows as $n^2$. Fixing $\beta_o$, say, the algorithm
directly estimates a partially summed density of states
\begin{equation}
\widehat{C}_{n,m_p}(\beta_o)=\sum\limits_{m_o}C_{n,m_p,m_o}e^{\beta_o m_o}\;.
\end{equation}
In this way, we can simulate lengths up to $n_{max}=1024$ at
fixed $\beta_o$.  
In a similar fashion, we also consider the diagonal
$\beta_p=\beta_o=\beta$, which is equivalent to considering the
partially summed density of states
\begin{equation}
\widetilde{C}_{n,m}=\sum\limits_{m_o+m_p=m}C_{n,m_p,m_o}\;.
\end{equation}
To reduce the error for our runs up to $n=1024$, we have taken
averages of ten independent runs each. Each run has taken
approximately 2 months on a 2.8GHz PC to complete.

\section{Results}
\label{results}

\begin{figure}[b!]
\includegraphics[width=7cm]{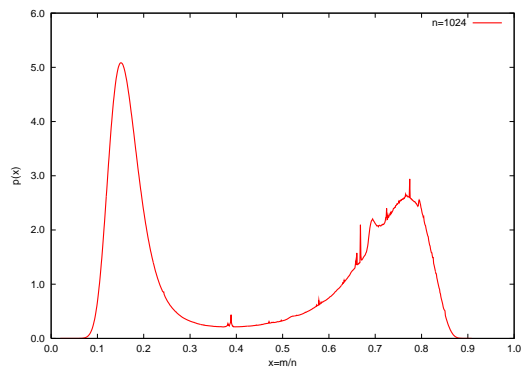}
\caption{\label{hist_2d}
Internal energy density distributions for the two-dimensional Foster-Seno model at the value of $\beta$ for
which the fluctuations are maximal, length $1024$.
}

\end{figure}
Before presenting the findings for our model, we briefly discuss the results of simulations
of the Foster-Seno model in two dimensions.
We find a first-order transition between a swollen coil and ordered
collapsed phase in agreement with Foster and Seno \cite{foster2001}.
Figure~\ref{hist_2d} shows the internal density distribution at
$\beta=\beta_c=1.04$, where the specific heat is maximal. This distribution is clearly
bimodal, and finite-size scaling supports the conclusion that the transition is
first-order. 
Our estimate of $\beta_c=1.04$ is close to the value $1.00(2)$ obtained by Foster and Seno
\cite{foster2001} from transfer matrix calculations. The low temperature phase is an ordered
$\beta$-sheet type phase.

\begin{figure}[ht]
\vspace*{-0.4cm}
\includegraphics[width=9cm]{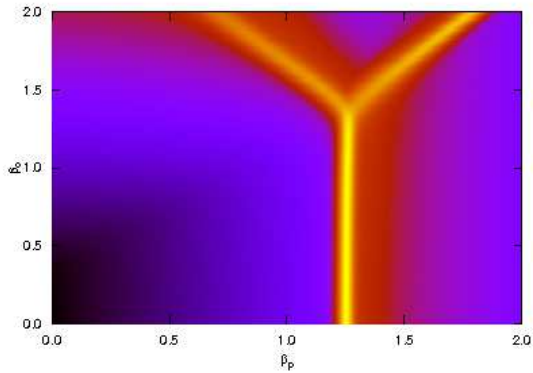}
\vspace*{-1.0cm}
\caption{\label{fl_3d}
This is a density plot of the logarithm of the largest eigenvalue of the matrix of
second derivatives of the free energy with respect to $\beta_p$
and $\beta_o$ at $n=128$. The lighter the shade the larger the value. 
}
\end{figure}

For the three-dimensional model, we have explored the full two-variable phase space
$(\beta_p,\beta_o)$ by using a two-parameter FlatPERM simulation 
of the model for lengths up to 128.
We performed $10$ independent simulations to ensure
convergence and understand the size of the statistical error in our
results. As in previous work \cite{krawczyk2005a,krawczyk2005b}, we
found the use of the largest eigenvalue of the matrix of second
derivatives of the free energy with respect to the parameters
$\beta_p$ and $\beta_o$ most advantageous to show the fluctuations in
a unified manner. 
Figure~\ref{fl_3d} displays a density plot of the size of fluctuations 
for $0\leq\beta_p,\beta_o\leq 2$. It suggests the presence of three 
thermodynamic phases separated by three phase transition lines meeting at
a single point. For small values of $\beta_p$ and $\beta_o$, we expect
the model to be in the excluded volume universality class of swollen polymers,
since at $\beta_p=\beta_o=0$ the model reduces to the simple self-avoiding walk.
The question arises as to the nature of the phases for large $\beta_p$ with $\beta_o$ fixed and
for large $\beta_o$ with $\beta_p$ small, and the type of transitions between each of the phases.

We find evidence
for a strong collapse phase transition when increasing $\beta_p$ for fixed
$\beta_o \lesssim 1.38$. Corrections to scaling at lengths $n\leq128$
make it difficult to
identify the nature of the transition. The location of the transition
seems independent of the value of $\beta_o$ and is located at
$\beta_p\approx 1.25$ for length $n=128$; this is taken from the location of
the peak of the fluctuations.  
Since our data indicate that this transition occurs for $\beta_o\lesssim 1.38$ at $\beta_p\approx 1.25$,
it follows that the diagonal line $\beta_o=\beta_p$ crosses this transition line.
Configurations in the collapsed phase are rich in parallel contacts; we shall discuss 
further details of the collapsed phase below.

\begin{figure}[ht]
\includegraphics[width=7cm]{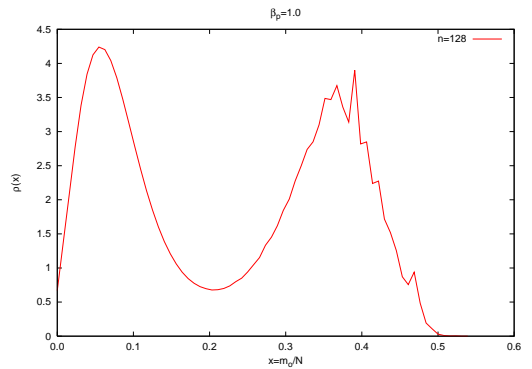}
\caption{\label{hist_3d_o}
A plot of the internal energy distribution in $m_o$ at $\beta_p=1.0$ for 
length $n=128$ at values of $\beta_o$ for which the fluctuations in $m_o$ are maximal.
}
\end{figure}

The situation changes significantly for  $\beta_o \gtrsim 1.38$.
When we start from the swollen phase at fixed $\beta_o > 1.38$ and increase 
$\beta_p$ we see evidence for a strong phase transition to a different
collapsed phase, in which orthogonal contacts are expected to play an important role.  
Further increase of $\beta_p$ leads to another strong transition to the parallel-contact rich phase.
We investigate the transition between the swollen coil and the
orthogonal-contact rich phase by considering the line $\beta_p=1.0$. 
Figure~\ref{hist_3d_o} shows a bimodal internal energy distribution at the
maximum of the fluctuations in $m_o$ for length $n=128$, indicating the presence of a first-order transition.

\begin{figure}[ht]
\includegraphics[width=9cm]{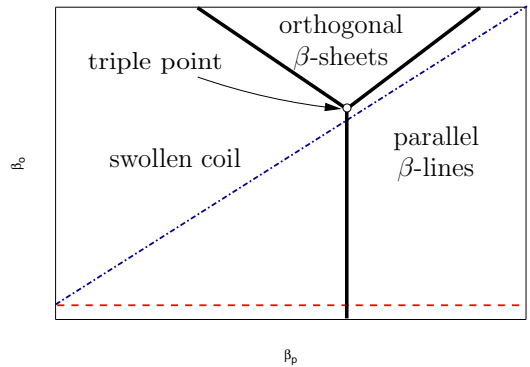}
\caption{\label{pd_3d}
  This figure represents our conjectured schematic phase diagram. The
  phases boundaries are marked by black lines.  
  The dotted-dashed (blue) and dashed (red) lines denote the lines
  along which we have performed one-parameter simulations.}
\end{figure}

Combining the evidence above, we conjecture
the phase diagram shown in Figure~\ref{pd_3d}, having
three phases and three transition lines that meet at a triple point
located at $(\beta^t_p,\beta^t_o)\approx(1.25,1.38)$ for length $n=128$. 
By considering the location of this point for different lengths $n$, 
we conclude that its estimate is affected by strong finite-size corrections to scaling.

\begin{figure}[ht]
\includegraphics[width=7cm]{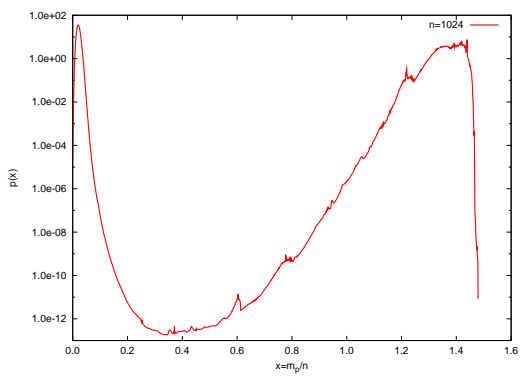}
\caption{\label{hist_p}
  Internal energy density distributions of $m_p$ at 
  $\beta_o=0$  and $\beta_p=0.996$ for length $1024$.}
\end{figure}

To further elucidate the nature of the phase transitions and the structure of the
low-temperature phases, we perform simulations for larger system sizes for the two
lines $\beta_o=0$ and $\beta_o=\beta_p$, using one-parameter FlatPERM simulations for 
lengths up to $n=1024$, averaged over ten independent simulations each.
We begin by considering $\beta_o=0$. The peak of
the specific heat occurs at $\beta_p=0.996$ for $n=1024$, which we
note is shifted away from the value at length $n=128$ and reflects the
presence of strong corrections to scaling. The distribution of $m_p$ at
this point is shown in Figure~\ref{hist_p}; we observe a clear bimodal distribution
with well-separated peaks and which ranges over fourteen orders of magnitude, 
convincingly supporting the conclusion of a first-order phase transition.
Similarly, along the line $\beta=\beta_o=\beta_p$ we find a single peak of
the specific heat, located at $\beta_p=0.998$ for $n=1024$. The distribution of
$m=m_o+m_p$ at this point 
displays the same
characteristics as the transition on the line $\beta_o=0$ described above.
Our investigations of the transition between 
the two collapsed phases were not conclusive, 
as it is difficult to do simulations at
very low temperatures. While we expect
there to be a first order phase transition between the two collapsed
phase we were unable to verify this. 

\begin{figure}[ht]
\vspace{-0.5cm}
\hfill
\includegraphics[width=4cm]{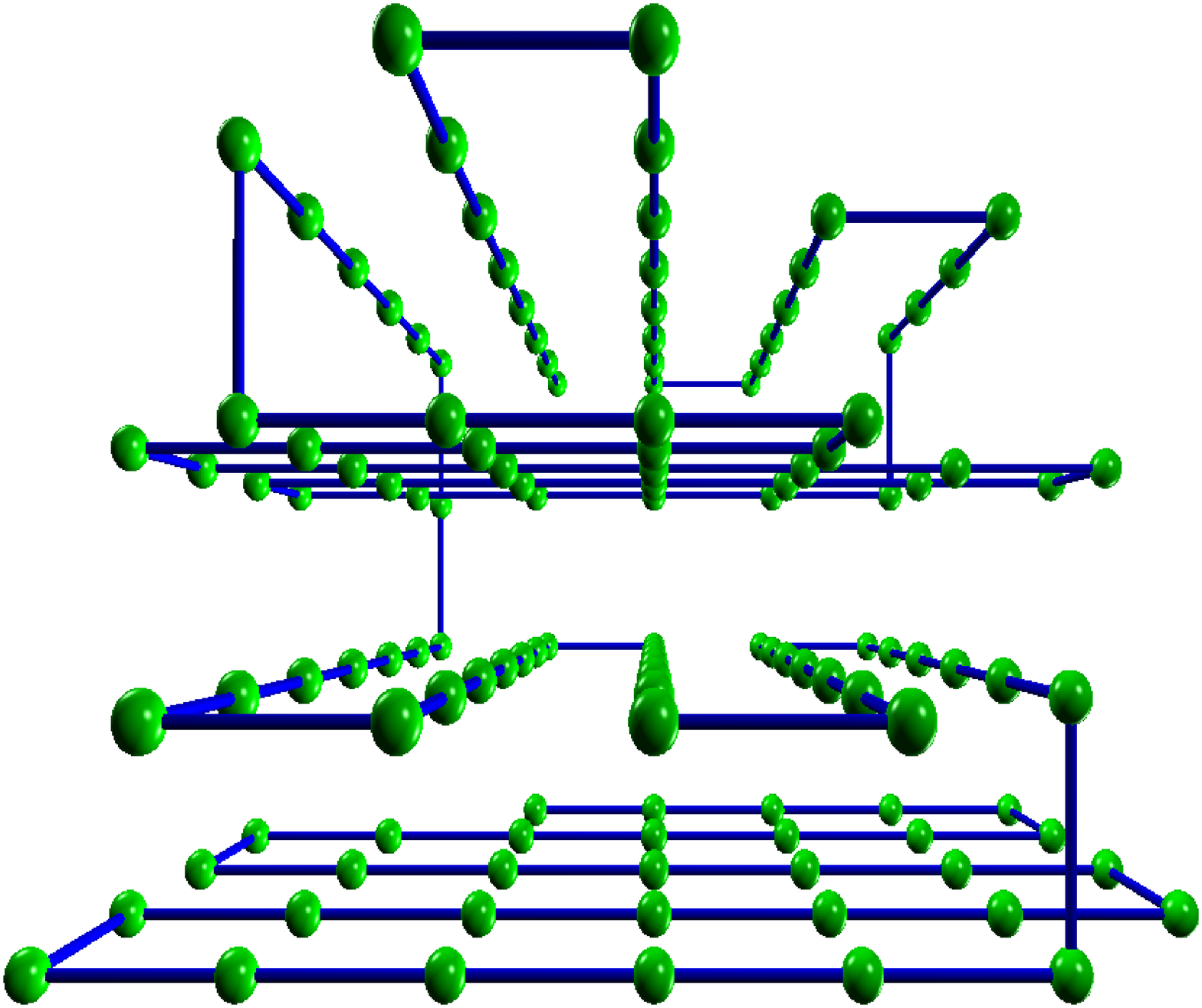}
\hfill
\includegraphics[width=4cm]{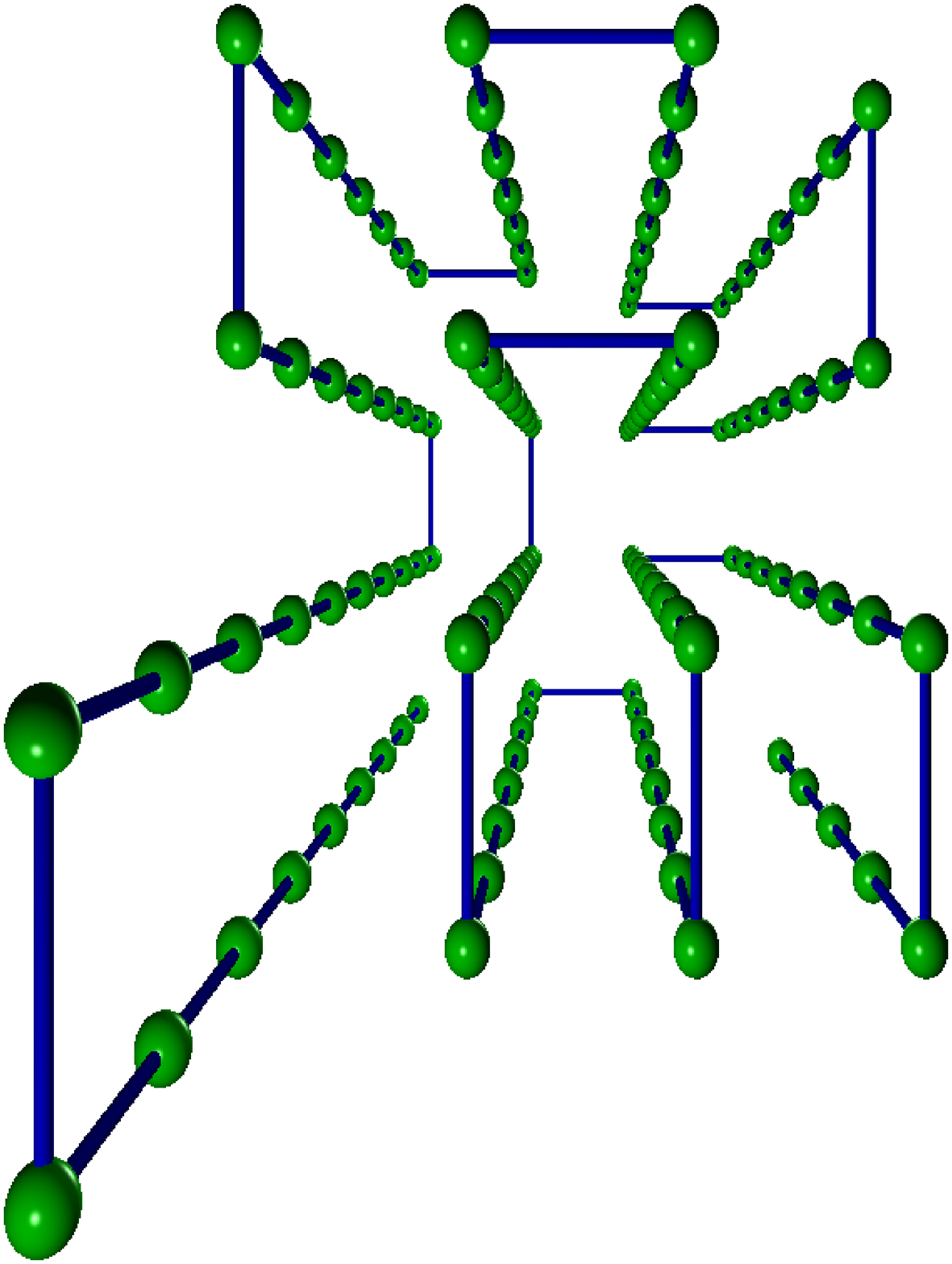}
\hfill
\caption{\label{configs}
Typical configurations for the two different collapsed phases, sampled at
$(\beta_p=1.3,\beta_o=1.9)$ (left) and at $(\beta_p=1.8,\beta_o=1.0)$ (right).}
\end{figure}

To delineate the nature of the two collapsed phases, we have randomly 
sampled typical configuration for each: two of these are shown in
Figure~\ref{configs}. In each case we have used $\beta_p>0$, where parallel
contacts are attractive.
For large $\beta_p$, we have a parallel contact rich phase, and typical
configurations have lines of monomers arranged in parallel. In Figure~\ref{configs},
there is a typical configuration for $(\beta_p=1.8,\beta_o=1.0)$, which demonstrates
these parallel $\beta$-lines.
For large $\beta_o$, orthogonal contacts
play an important role. A typical configuration for $(\beta_p=1.3,\beta_o=1.9)$
consists of parallel lines arranged in $\beta$-sheets, which are layered orthogonally.
The entropy of the phase consisting out of orthogonal $\beta$-sheets is lower than
the entropy of the phase consisting out of collection of parallel lines, which explains
why the collapse-collapse transition line is shifted away from the diagonal.
Clearly the formation of $\beta$-sheets is dependent on $\beta_p$
being positive (attractive parallel) interactions. 

In conclusion, we have demonstrated the intriguing possibility of
building a secondary structure in proteins, which involves layered
$\beta$-sheets structures interacting in two different ways. Depending
on the modelling of the interactions we distinguish $\beta$-sheets
that align parallel or orthogonal to each other, which leads to two
different phases, and secondary structures.  There remains an
interesting theoretical question as to the behaviour of the system
when the parallel interactions are repulsive but the orthogonal
interactions are highly attractive.
 
 Financial support from the Australian Research Council and the Centre
 of Excellence for Mathematics and Statistics of Complex Systems is
 gratefully acknowledged by the authors.

\end{document}